\documentclass{resonance}
\usepackage{amsfonts}
\usepackage{amsmath}
\usepackage{amssymb}
\usepackage{rotating}
\usepackage{epsfig}  
\usepackage{graphics} 
\usepackage{graphicx} 
\usepackage{txfonts}
\usepackage{marginnote}
\usepackage{array}
\usepackage{geometry}
\usepackage{color}      
\definecolor{light}{rgb}{0.8,0.5,0.5}
\definecolor{cblue}{rgb}{0.9,0.9,1.0}
\definecolor{darkblue}{rgb}{0.1,0.1,0.6}
\definecolor{darkred}{rgb}{0.6,0.1,0.1}
\usepackage{cleveref}
\crefformat{section}{\S#2#1#3}
\crefformat{subsection}{\S#2#1#3}
\crefformat{subsubsection}{\S#2#1#3}
\crefrangeformat{section}{\S\S#3#1#4 to~#5#2#6}
\crefmultiformat{section}{\S\S#2#1#3}{ and~#2#1#3}{, #2#1#3}{ and~#2#1#3}
\newcommand{\bed}{\begin{displaymath}}
\newcommand{\eed}{\end{displaymath}}
\newcommand{\bei}{\begin{itemize}}
\newcommand{\eei}{\end{itemize}}
\newcommand{\bef}{\begin{figure}}
\newcommand{\eef}{\end{figure}}
\newcommand{\ben}{\begin{enumerate}}
\newcommand{\een}{\end{enumerate}}
\newcommand{\beq}{\begin{equation}}
\newcommand{\eeq}{\end{equation}}
\newcommand{\ber}{\begin{eqnarray}}
\newcommand{\eer}{\end{eqnarray}}

\newcommand{\lsim}{\raisebox{-0.3ex}{\mbox{$\stackrel{<}{_\sim} \,$}}}
\newcommand{\gsim}{\raisebox{-0.3ex}{\mbox{$\stackrel{>}{_\sim} \,$}}}

\newcounter{attnctr} 

\begin{document}

\title{The Sounds of Music : Science of Musical Scales}
\secondTitle{I : Human Perception of Sound}
\author{Sushan Konar}

\maketitle
\authorIntro{\includegraphics[width=2.5cm]{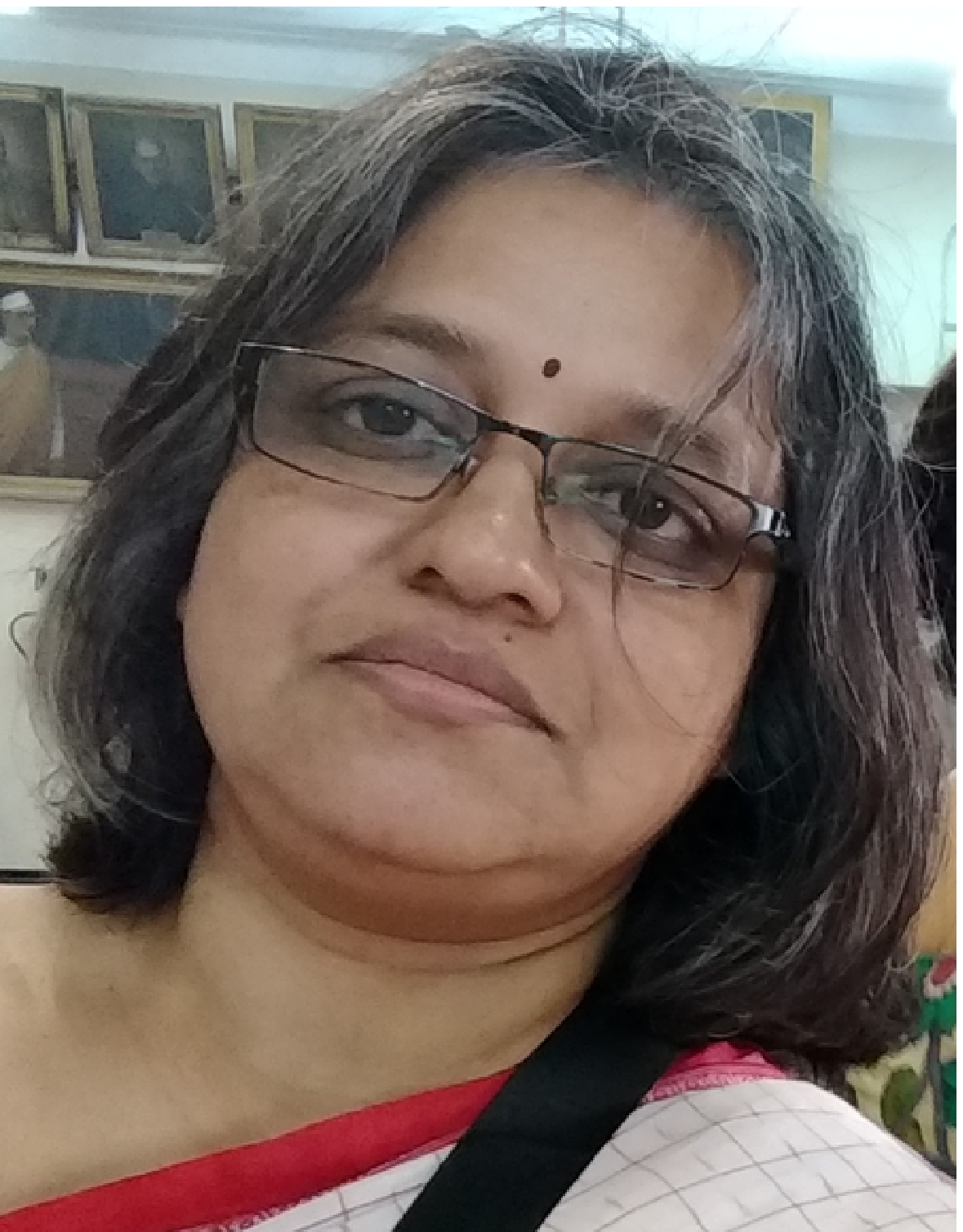}\\
  Sushan Konar  works  on  stellar  compact objects. She also writes popular
  science articles and maintains a weekly astrophysics-related blog called
  `{\em Monday Musings}'.}
\begin{abstract}
Both, human appreciation  of music and musical  genres, transcend time
and space.  The universality of  musical genres and associated musical
scales is  intimately linked to the  physics of sound and  the special
characteristics  of human  acoustic  sensitivity.  In  this series  of
articles, we  examine the  science underlying  the development  of the
heptatonic  scale, one  of the  most  prevalent scales  of the  modern
musical genres, both western and Indian.
\end{abstract}
\monthyear{2018}
\artNature{GENERAL  ARTICLE}


\section*{Introduction}
\label{s-intro}
Fossil records  indicate that the  appreciation of music goes  back to
the dawn  of human  sentience and  some of the  musical scales  in use
today could also  be as ancient.  This universality  of musical scales
likely owes its  existence to an amazing  circularity (or periodicity)
inherent  in  human sensitivity  to  sound  frequencies. Most  musical
scales are  specific to a particular  genre of music and  there exists
quite a number  of them. However, the `heptatonic'  (having seven base
notes) scale happen  to have a dominating presence in  the world music
scene today.  It  is interesting to see  how this has more  to do with
the physics of  sound and the physiology of  human auditory perception
than history.  We shall  devote this  first article  in the  series to
understand the specialities of human response to acoustic frequencies.


\begin{figure}
\caption{Range of acoustic (sonic)  frequencies. Frequencies above and
  below  this   range  are  known   as  {\em  ultra-sonic}   and  {\em
    infra-sonic} frequencies respectively.}
\label{f-range}
\vspace{-0.5cm}
\begin{center}
\centering\includegraphics[width=12.0cm]{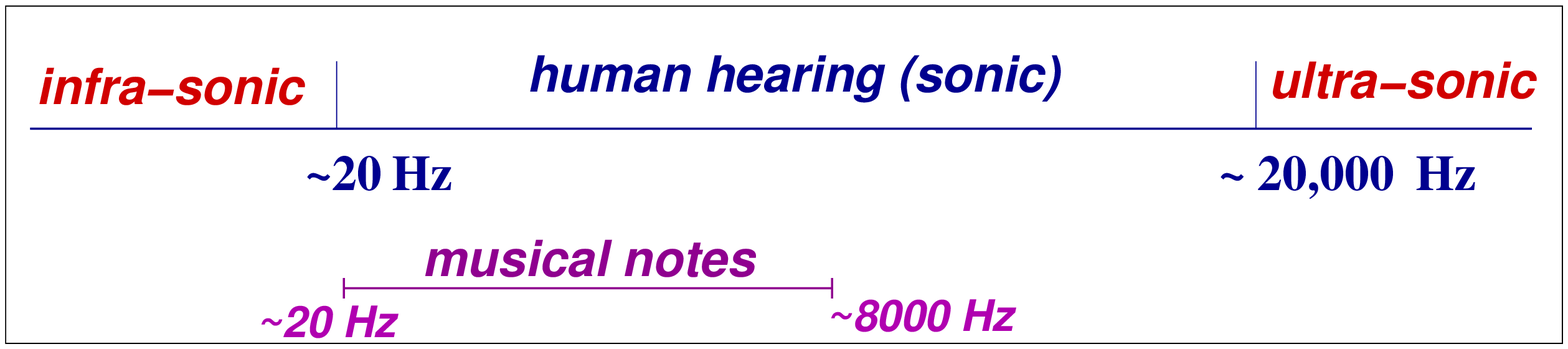}
\end{center}
\vspace{-0.5cm}
\eef

Human ear  is a remarkable organ  in many ways.  The  range of hearing
spans  three   orders  of  magnitude  in   frequency,  extending  from
$\sim$20~Hz  to $\sim$20,000~Hz  even  though  the sensitivity  varies
widely over this  range.  Unlike the electromagnetic,  sound waves are
longitudinal  pressure  waves.   The   regions  of  compression  (high
pressure) and  rarefaction (low pressure)  move through a  medium with
the  speed of  sound  (appropriate for  that  medium). The  separation
between  pressure peaks  (or troughs)  corresponds to  one wavelength,
appropriate  for  the specific  frequency.   The  modification to  the
ambient pressure, induced by the  propagation of such a pressure wave,
is sensed  by the human ear  and is interpreted as  sound. The minimum
pressure difference  to which the  ear is  sensitive is less  than one
billionth ($> 10^{-9}$)  of one atmosphere, meaning the  ear can sense
(hear) a wave for which the  pressure at the peaks (or troughs) differ
from  the ambient  atmospheric  pressure  by less  than  1  part in  a
billion.  Evidently,  human ear  is an extremely  sensitive instrument
even though the  sensitivity varies quite a lot from  person to person
and is not constant over the entire audible frequency range.
\begin{table}[h]
\caption{Sensitivity characteristics of human auditory system. }
\label{t-ear}
\centering{
\begin{tabular}{|ll|} \hline
\multicolumn{2}{|c|}{} \\
\multicolumn{2}{|c|}{Human Ear} \\
directional coverage                   & 360$^\circ$ \\  
location accuracy                      & $\sim 5^\circ$ \\  
frequency coverage (maximum/minimum)   & $ \sim 10^3$ \\  
intensity resolution (maximum/minimum) &  $\sim 10^{9} : 1$ \\
& \\ \hline
\end{tabular}
}
\vspace{-0.5cm}
\end{table}

Taken together, the  basic characteristics of our ears are  as seen in
Table[\ref{t-ear}]. For ordinary human communication, a frequency span
of a  decade (ratio of  highest to lowest  frequency $\sim$ 10)  and a
pressure sensitivity  ratio of  10$^4$:1 would suffice.  The extremely
large range  of human auditory  sensitivity strongly imply  that there
must be some other purpose that the human ear is supposed to serve. It
is suspected that the ear  may have primarily evolved for self-defence
(as  human  beings  are  endowed  with  very  little  in  the  way  of
self-defence compared  to other big animals).   Language and enjoyment
of music are likely to be evolutionary by-products.

\section{Sound Waves}
\label{s-sound}
\subsection{Standing Waves}
\label{ss-stand}
It appears  that the modern Western  musical scale has its  origins in
the  tuning of  a harp-like  instrument called  the `lyre'  of ancient
Greece.   This  instrument was  a  `tetra-chord'  (consisting of  four
strings tied  at both ends) and  these used to be  `plucked' to create
music. The vibrations,  thus generated on a string tied  at both ends,
are {\em standing}  or {\em stationary} waves as they  appear to stand
still  instead  of travelling.   In  a  1-dimensional medium,  like  a
string,  a standing  wave is  produced when  two waves  with the  same
frequency (therefore, of same  wavelength) and amplitude travelling in
opposite  directions interfere.   Such  a standing  wave is  therefore
generated  in a  string  by  the initial  `pluck',  which induces  two
identical waves  that travel in  opposite directions and  reflect back
and forth  between the fixed  boundary points  at two ends.   The peak
amplitude of a  standing wave at any point in  space (along the length
of  the string,  in  this  case) remains  constant  in  time, and  the
oscillations at different points throughout the wave stay in phase.

We can express a pair of  harmonic waves travelling along the positive
and the negative $x$-axis as -
\ber
f_1(x,t) &=& A_0 \sin({2\pi x \over \lambda } - 2 \pi \nu t) \,,  \\
f_2(x,t) &=& A_0 \sin({2\pi x \over \lambda } + 2 \pi \nu t) \,; 
\eer
where, $x$  and $t$ are position  and time co-ordinates; $A_0$  is the
amplitude and  $\nu$ is  the frequency of  the wave.   The wavelength,
$\lambda$, is equal to  $u/\nu$ where $u$ is the speed  of the wave in
the medium of propagation. Therefore, the resultant wave, obtained from
a superposition of these two are given by, 
\ber
f(x,t) &=& f_{1} + f_{2} \nonumber \\
       &=& A_0 \sin({2\pi x \over \lambda } - 2 \pi \nu t) 
           + A_0 \sin({2\pi x \over \lambda } + 2 \pi \nu t) \nonumber \\ 
       &=& 2 A_0 \sin \left({2\pi x \over \lambda }\right) \cos(2 \pi \nu t) \,. 
\eer
As can be seen, the amplitude of this wave at a location $x$, given by
$2A\sin  \left({2\pi   x  \over   \lambda  }\right)$,   is  stationary
(time-independent). Hence,  the name,  standing wave.  It can  also be
seen from this that the amplitude goes to zero when -
\beq
x=\;...-{3\lambda     \over    2},\;-\lambda     ,\;-{\lambda    \over
  2},\;0,\;{\lambda \over 2},\;\lambda ,\;{3\lambda \over 2},.. \,,
\eeq
and reaches the maximum value when
\beq
x=\;...-{5\lambda  \over  4},\;-{3\lambda \over  4},\;-{\lambda  \over
  4},\;{\lambda  \over   4},\;{3\lambda  \over   4},\;{5\lambda  \over
  4},..\,.
\eeq
These points are called `node's and `anti-node's respectively.

\begin{figure}[h]
\begin{center}
\centering\includegraphics[width=15.0cm]{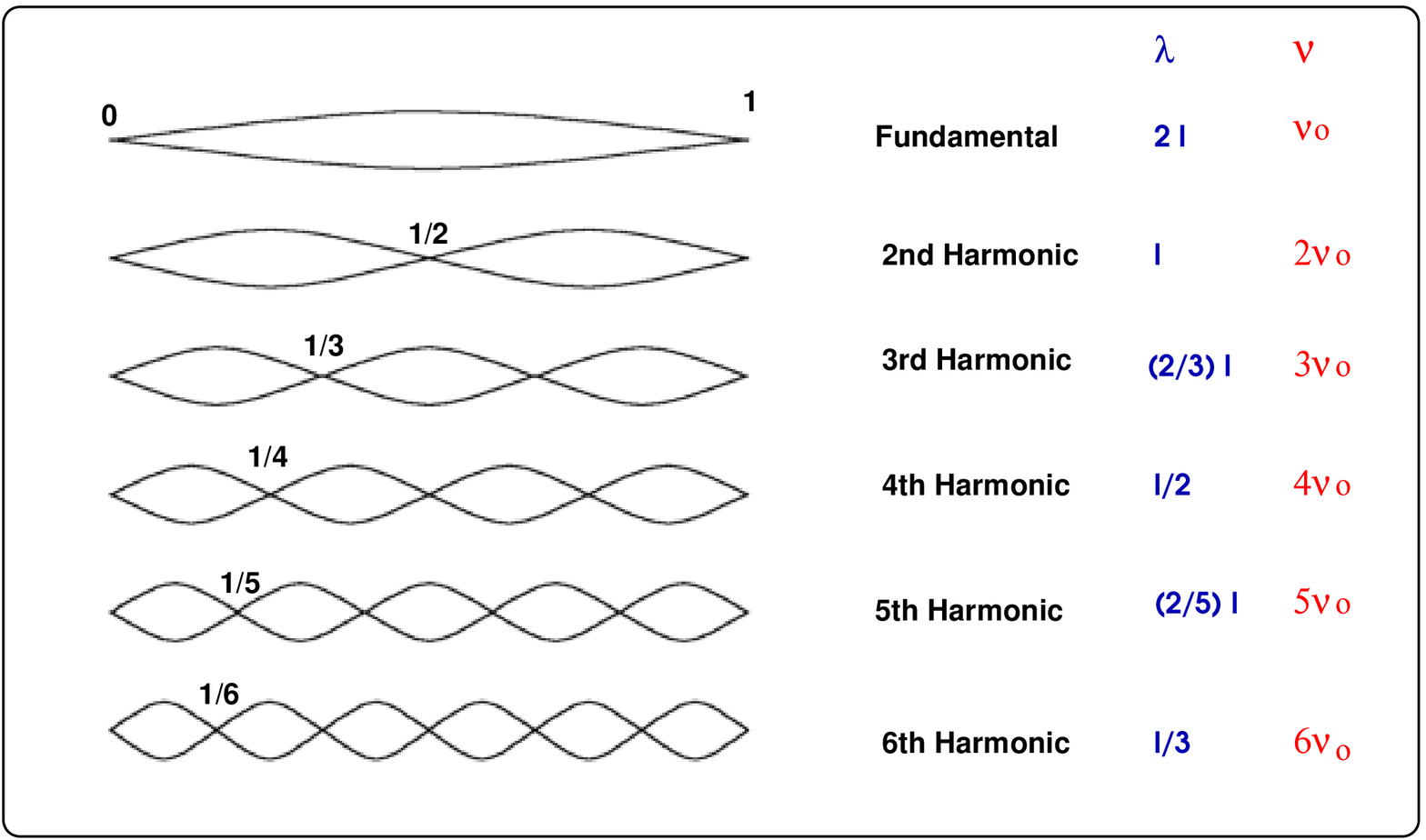}
\end{center}
\caption{Vibrational modes of a plucked  string of length {\bf l}. The
  wavelengths  and the  corresponding frequencies  of the  fundamental
  ($\nu_o$) and some of the higher harmonics have been illustrated.}
\label{f-string}
\vspace{-0.75cm}
\eef

Now,  for  a  string  of  length  {\bf l}  fixed  at  both  ends,  the
fundamental frequency is given by the vibration whose nodes are at the
two ends of  the string. Therefore, the wavelength  of the fundamental
frequency is {\bf 2l}. The fundamental frequency is then given by,
\beq
\nu = \frac{u}{2L} \,,
\label{e-funda}
\eeq
where $u$ is the velocity of wave propagation which can be obtained as,
\beq
u = \sqrt{\frac{T}{\mu}} \,.
\eeq
where $T$ is  the tension in the  string and $\mu$ is  the linear mass
density of the string. This implies  that - a) the shorter the string,
the  higher  the frequency  of  the  fundamental,  b) the  higher  the
tension, the higher  the frequency of the fundamental,  c) the lighter
the string, the higher the frequency of the fundamental. Therefore, it
is possible to modify the fundamental frequency of a string by varying
any  one of  these three  quantities  (length, mass  per unit  length,
tension).   Moreover,  the  $n$-th   harmonic  is  obtained  when  the
wavelength  is $\lambda_{n}=2L/n$.  Therefore,  the  frequency of  the
$n$-th harmonic is obtained to be,
\beq
\nu_{n} = \frac{nu}{2L} \, = \frac{n}{2L} \sqrt{\frac{T}{\mu}} \,, 
\label{e-nu}
\eeq
as  seen  in Fig.[\ref{f-string}].  It  is  obvious  that the  set  of
possible vibrational modes supported by  such a string is quantised in
nature, as  only a discrete  set of possible wavelengths  are allowed.
This  quantisation property  comes  from the  boundary condition  that
requires the string to have zero vibrational amplitude at each end.

\subsection{Beat Frequency}  
\label{ss-beat}
Once the sound is produced (for example in the string discussed above,
or by the  human voice or by  any other agent) it  travels through the
intervening medium  (the air surrounding  us, in most cases)  to reach
the human  ear. When two  such waves,  nearly equal in  frequency, are
sounded together they produce beats.

\begin{figure}[h]
\begin{center}
\centering\includegraphics[width=12.5cm]{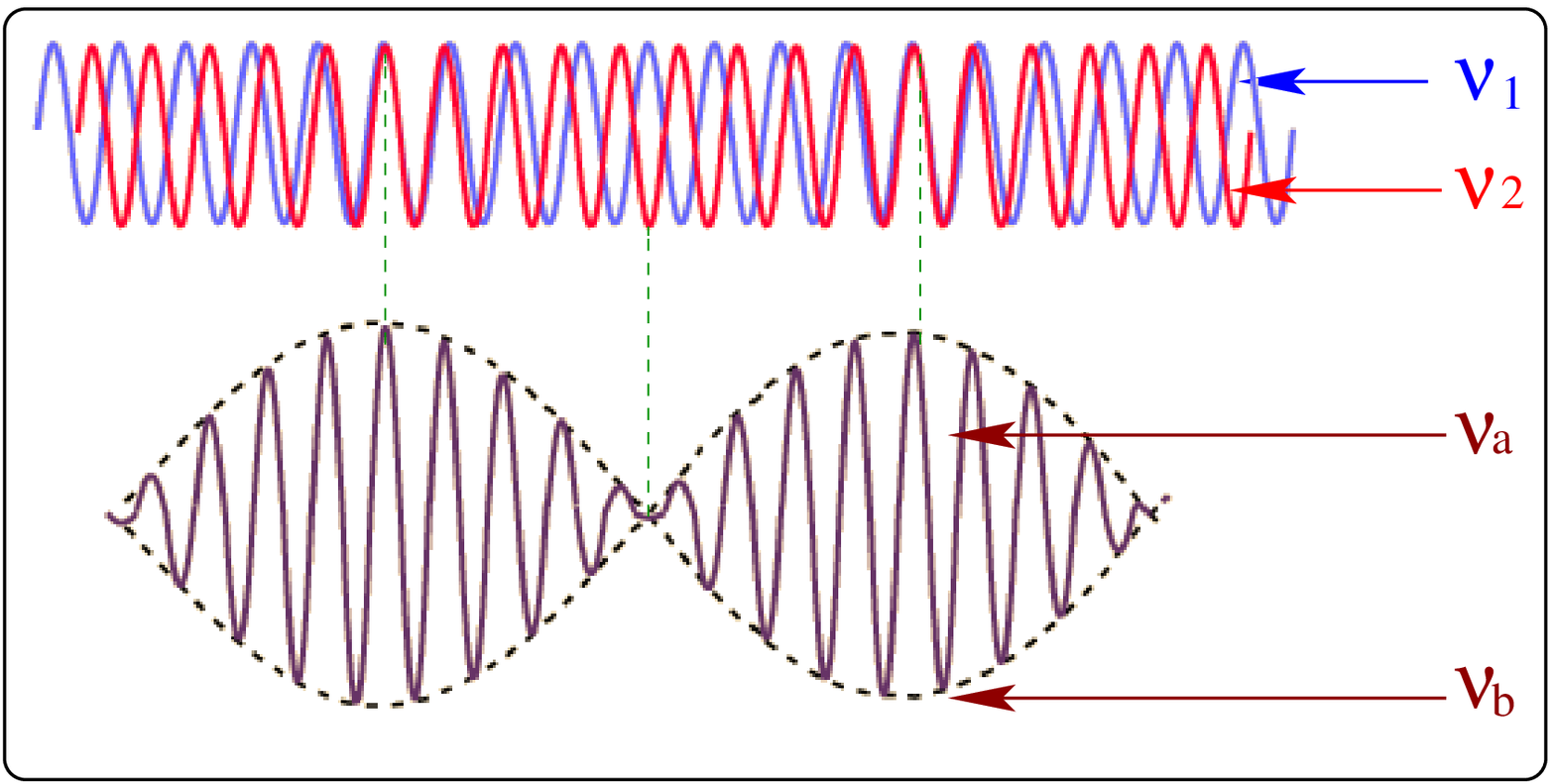}
\end{center}
\caption{Superposition  of  two  waves with  frequencies  $\nu_1$  and
  $\nu_2$.  The  resultant is a  wave of frequency $\nu_a$  ($=(\nu_1 +
  \nu_2)/2$)  modulated by  a wave of frequency  $\nu_b$ ($=(\nu_1  -
  \nu_2)/2$).}
\label{f-beat}
\vspace{-0.75cm}
\eef

Let us consider two such travelling waves with frequencies $\nu_1$ and
$\nu_2$ and equal amplitude. The superposition of these is given by,
\ber
F(t) = f_1(t) + f_2(t)
     &=& A \cos(2 \pi \nu_1 t) + A \cos(2 \pi \nu_2 t)  \nonumber \\
     &=& 2 A \cos(\pi (\nu_1 + \nu_2) t) \, \cos(\pi (\nu_1 - \nu_2) t)  \,.
\eer
Clearly,  these appear  as two  correlated waves,  one modulating  (or
acting as an `envelope' for) the  other; where one of the frequencies,
$\nu_a$, is equal to the  average of the original frequencies ($(\nu_1
+  \nu_2)/2$)  and  the  other,  $\nu_b$, is  equal  to  half  of  the
difference ($(\nu_1 - \nu_2)/2$).  When $\cos (\pi (\nu_{1} - \nu_2)t)
=  1$,  the  two   (original)  waves  are   in  phase   and  interfere
constructively. When this quantity vanishes the waves are out of phase
and interfere destructively.

It should be noted that, in the modulation pattern, every second burst
in the  modulation pattern  is inverted.  Each peak  is replaced  by a
trough and  vice-versa. Since  the human ear  is not sensitive  to the
phase of a sound, only to its amplitude or intensity, the frequency of
the envelope  appears to  have twice the  frequency of  the modulating
wave.  Therefore, the audible beat frequency is given by
\beq
\nu_{\rm beat} = 2 \nu_b = \nu_{1} - \nu_{2} \,.
\eeq

\section{Human Auditory Perception}
\label{s-human}
Whatever may have been the evolutionary  logic, the human ear has been
endowed with  excellent sensitivity  to sound. This  sensitivity comes
accompanied with  an appreciation  for harmonic  relationships between
the frequencies present in a given  sound. This, in turn, manifests as
two special characteristics of human hearing, as described below.

{\em \bf Consonance \& Dissonace :} Human perception of harmony between
two  tones (a  {\em tone}  is the  sound corresponding  to one  single
frequency) is surprisingly arithmetic, meaning that the closer the two
tones  are  to  having  a   simple  ratio  of  frequencies,  the  more
``harmonious" (or consonant) they sound.  This happens because, in any
musical  instrument,  instead of  a  pure  tone  we generally  have  a
harmonic series comprising of the fundamental frequency and its higher
harmonics.   In general,  the  fundamental frequency  (say, $\nu$)  is
expected  to  appear  with  the maximum  amplitude  while  the  higher
harmonics  ($2\nu, 3\nu,  4\nu$..)   would  appear with  progressively
smaller amplitudes.  In presence of these harmonics the notes (musical
frequency) which  differ by a  ratio of $a/b$  will share many  of the
early (and  therefore stronger)  harmonics, when $a$  and $b$  are two
small integers.  For each integer $n$ the $nb$-th harmonic of one will
be the $na$-th harmonic of the other. That is,
\beq
\nu_1 =  \frac{a}{b} \, \nu_2  \; \;
\Rightarrow \; \;  nb \, \nu_1  = na \, \nu_2 \,,
\label{e-harmonic}
\eeq
implying that tones  whose fundamentals are $\nu_1$  and $\nu_2$ would
sound sweet, or consonant, together.

Recent  studies have  shown  that  the human  brain  has two  separate
centres for processing consonant and dissonant (not harmonious) sounds
which  are associated  with  different emotions.  This `explains'  why
humans  associate  `pleasure'  and `displeasure'  with  consonant  and
dissonant  sounds.   Interestingly,  natural sounds  evoking  negative
emotional  responses,   such  as   -  avalanches,   gales/high  winds,
tornadoes, or human/animal sounds of somewhat negative nature (groans,
shrieks, howls,  roars) - typically have  non-integer-related harmonic
content; whereas our own human singing voices are inherently consonant
in  nature  (i.e.   have   integer-related  harmonic  content).

{\em \bf  Tone vs.  Pitch  :} Human beings can  hear a large  range of
acoustic  frequencies,  but our  perception  of  sound has  a  curious
periodicity  to it.   Therein  lies the  concept of  {\em  pitch} -  a
subjective,  perceived (by  humans) aspect  of sound,  associated with
musical tones. A musical tone is  defined by its frequency.  The pitch
of  a  tone is  closely  related  to  this  frequency.  In  fact,  the
frequency and the pitch of a  tone can effectively be thought as being
`proportional' (but not quite).  Let us consider two pure tones having
the same frequency $\nu$. When these two are heard together, the sound
is said  to be  in `unison'  - both the  notes are  the same.   Let us
increase one  of the frequencies,  such that  there is now  a non-zero
difference  in  the  frequencies  ($\delta \nu  \neq  0$).   From  the
discussion in the previous section, it  can be seen that now we should
have a  sound wave of  base frequency equal to  $\nu_a = \nu  + \delta
\nu/2$, modulated by a beat frequency of $\delta \nu$.  Evidently, the
base  frequency is  very close  to  the original  frequency for  small
values of $\delta \nu$.  For $\delta  \nu \lsim 12$~Hz, human ear only
detects the  sound combination as  a single  tone (known as  a `fused'
tone), because the amplitude modulation for such cases are too slow to
be detectable by human auditory system.

When $\delta  \nu \gsim 12$~Hz, the  human ear begins to  perceive two
separate tones with some  roughness (like `buzzing').  This perception
continues till  the difference is  large enough  for the human  ear to
perceive two separate  tones with clarity.  It needs to  be noted that
the value of $\delta \nu$, at  which this clarity is achieved, depends
strongly on  $\nu_a$.  For audio  frequencies near the upper  limit of
human hearing, $\delta \nu$ could  be as large  as 400~Hz for  the two
notes to be perceived separately.

If we  now continue  to increase the  frequency difference,  human ear
would continue  to perceive the two  tones as two separate  ones - but
only up-to a point. As the  second tone approaches twice the frequency
of the  first, its `pitch' returns  to its starting point.  This gives
rise to the  situation when two tones, whose frequencies  are a factor
of two  apart, are perceived to  be the same musical  `note', although
their frequencies are  different.  These two notes are said  to be one
{\bf \em octave}  apart in Western musical tradition,  an octave being
the interval between one musical note  and another with half or double
its  frequency.    In  frequency  space,  the   interval  between  the
fundamental and second harmonic  (or between any successive harmonics)
of a harmonic series, is an octave.

All  the notes  in between  unison  and the  octave are  said to  have
different {\em chroma}, or colour, and  notes with the same chroma but
separated  by one  or more  octaves are  said to  have different  tone
height.  While chroma applies to  a continuous variation in frequency,
pitch refers  to tones with  specifically defined values  of frequency
modulo an integral  multiple of octaves.  This perception  of a unique
note across octaves is a natural  phenomenon that has been referred to
as the `basic  miracle of music', the  use of which is  common in most
musical systems.  Because  the notes that are an  integral multiple of
octave apart  `sound' same, they  are given  the same name  in musical
notations.  This is called the {\bf \em octave equivalence} of musical
frequencies (the  assumption being  that the  frequencies one  or more
octaves  apart  are  musically  equivalent)  and is  a  part  of  most
contemporary musical cultures.

Other similar relations  have been seen to exist which  have also been
of great importance to our  musical tradition.  For example, two notes
which differ in frequency by a factor of 3/2 sound extremely consonant
together and  this relation  is known  as the  {\em consonance  of the
  fifth}. We  shall see that  this relation  (or the more  general one
described by  Eq.[\ref{e-harmonic}]) is actually a  consequence of the
octave equivalence,  which can be  understood in terms  of combination
tones (or beat frequencies).

A musical fifth comprises of two basic frequencies $\nu_1$ and $\nu_2$
where $\nu_2 =  3/2 \nu_1$.  The superposition of  these two generates
the beat frequency $\nu_b = \nu_2 - \nu_1 = \nu_1/2$, which is exactly
one octave below $\nu_1$. Therefore,  the tones generated by a musical
fifth ($\nu_1/2, \nu_1,  3/2 \nu_1$) constitute the  first three terms
in a harmonic series, where  $\nu_b$ is the `fundamental', and $\nu_1$
and $\nu_2$  are the second  and the  third harmonics.  Thus  when two
tones  constituting a  musical fifth  are played,  the periodicity  of
their superposition  curve enables  the auditory  system to  sense the
presence of  a fundamental tone that  is not actually present,  and is
therefore called a  `missing fundamental'.  In other  words, a musical
fifth is perceived as a harmonic  series by the human auditory system.
Similarly,  for a  sequence of  harmonics consisting  of even  and odd
integral  multiples  of  the  fundamental,  each  pair  of  successive
harmonics create a combination tone  with periodicity equal to that of
the  fundamental   (which  is   quite  remarkable).    Therefore,  the
fundamental  is perceived  even when  it is  absent from  the original
sequence.

This particular ability of human auditory system to identify consonant
sound and  sense the  {\em pitch} rather  than the  absolute frequency
(tone) of  a note is  actually at the root  of our musical  scales. We
shall see in the next article how it gave rise to our familiar western
classical scale of music.


\begin{thebibliography}{99} 
\bibitem{berg95}
  R.~E. Berg and D.~G. Stork,
  \textit{The Physics of Sound},
  Prentice Hall, New Jersey, 1995.
\bibitem{benso06} 
  D. Benson,
  \textit{Music: A Mathematical Offering},
  Cambridge University Press, 2006.
\bibitem{elmor69}
  W.~C. Elmore and N.~A. Heald,
  \textit{Physics of Waves},
  McGraw-Hill, 1969.
\bibitem{gill09}
  K.~Z. Gill and D. Purves,
  \textit{A Biological Rationale for Musical Scales},
  PLOS One, {\bf 4(12)}, e5144, 2009.
\bibitem{schel94}
  E.~G. Schellenberg and S.~E.Trehub,
  \textit{Frequency ratios and the perception of tone patterns},
  Psychonomic Bulletin \& Review, {\bf 1(2)}, pp.191-201, 1994.
\end{thebibliography}
\end{document}